\documentclass[aps,prl,preprint,superscriptaddress,showpacs]{revtex4}

\usepackage{amsmath}
\usepackage{graphicx}

\begin{document}

\title{Fundamental Behavior of Electric Field Enhancements in the Gaps Between Closely Spaced Nanostructures}

% Authors / affiliations
\author{Jeffrey M. McMahon}
\affiliation{Department of Chemistry, Northwestern University, Evanston, IL 60208}
\affiliation{Center for Nanoscale Materials, Argonne National Laboratory, Argonne, IL 60439}
%\email[]{jeffrey-mcmahon@northwestern.edu}

\author{Stephen K. Gray}
\affiliation{Center for Nanoscale Materials, Argonne National Laboratory, Argonne, IL 60439}

\author{George C. Schatz}
\affiliation{Department of Chemistry, Northwestern University, Evanston, IL 60208}
\email[]{schatz@chem.northwestern.edu}

% Date
\date{\today}

% Abstract
\begin{abstract}
We demonstrate that the electric field enhancement that occurs in a gap between two closely spaced nanostructures, such as metallic nanoparticles, is the result of a transverse electromagnetic waveguide mode. We derive an explicit semianalytic equation for the enhancement as a function of gap size, which we show has a universal qualitative behavior in that it applies irrespective of the material or geometry of the nanostructures and even in the presence of surface plasmons. Examples of perfect electrically conducting and Ag thin-wire antennas and a dimer of Ag spheres are presented and discussed.
\end{abstract}


% 03.50.De == Classical electromagnetism, Maxwell equations
% 71.45.Lr == Charge-density-wave systems

% 78.67.-n == Optical properties of low-dimensional, mesoscopic, and nanoscale materials and structures
% 62.23.Hj == Nanowires

% 42.25.Fx == Diffraction and scattering
% 42.25.Gy == Edge and boundary effects; reflection and refraction
% ******************************

%\keywords{}

\maketitle


Structures that generate large electric field enhancements relative to the incident field, hereon referred to as $|\textbf{E}|^2$ enhancements, have recently received a great deal of attention \cite{SERS_review_Moskovits-RevModPhys1985, rough-surf_SERS_theory_Pendry-PRL1996, bowtie_antennas_Moerner-PRL2005, opt-antennas_OJFMartin_Science-2005, EM-enh_capacitor_model_Park-PRL2009, McMahon_2010_JACS_SERS-trimers, EM_fields_NP-dimers_GCS, coupled-nanorods_Garcia-de-Abajo_PRB-2005, coupled-NPs_small-gaps_Garcia-de-Abajo_OptExp-2006, EM_fields_dimers_Martin-2001}. This is because such enhancements are central to a number of physical processes, including such surface-enhanced spectroscopy techniques as surface-enhanced Raman scattering (SERS), second harmonic generation, and enhanced absorption and fluorescence \cite{SERS_review_Moskovits-RevModPhys1985}. Often the enhanced field is generated in the small crevices of a roughened metal surface \cite{rough-surf_SERS_theory_Pendry-PRL1996} or at the junctions of closely spaced nanoparticles \cite{bowtie_antennas_Moerner-PRL2005, opt-antennas_OJFMartin_Science-2005, EM-enh_capacitor_model_Park-PRL2009, McMahon_2010_JACS_SERS-trimers, EM_fields_NP-dimers_GCS, coupled-nanorods_Garcia-de-Abajo_PRB-2005, coupled-NPs_small-gaps_Garcia-de-Abajo_OptExp-2006, EM_fields_dimers_Martin-2001}. Herein we focus on the latter structures, and while much is known about the $|\textbf{E}|^2$ enhancements in them, there is still confusion over fundamental principles. In particular, the functional dependence on gap size \cite{bowtie_antennas_Moerner-PRL2005, EM-enh_capacitor_model_Park-PRL2009, coupled-nanorods_Garcia-de-Abajo_PRB-2005, coupled-NPs_small-gaps_Garcia-de-Abajo_OptExp-2006}, arguably the most basic and important aspect, has not been quantitatively determined and the underlying physical principles which determine it are not entirely known. It is the purpose of this Letter to resolve this issue through finite element method (FEM) calculations \cite{Jin_FEM} and an analytical theory developed for the transmission of light through an isolated slit in a metal film \cite{modal-analysis_FJGV_2004}.

\begin{figure}
  \includegraphics[scale=0.36, bb=0 0 669 244]{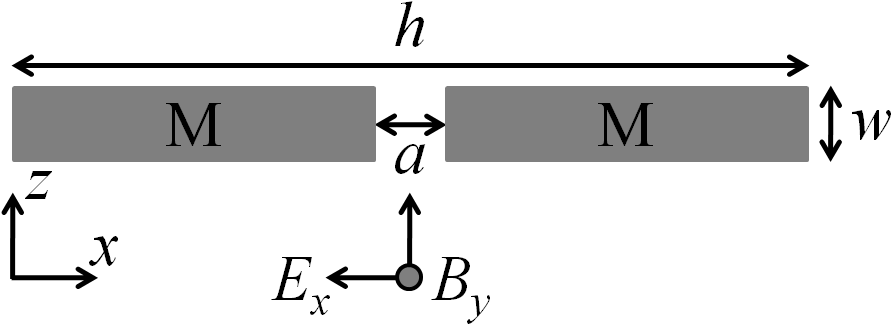}
  % schematic_small-w.png: 892x325 pixel, 96dpi, 23.60x8.60 cm, bb=0 0 669 244
  \caption{Schematic diagram of a thin-wire antenna. The parameters shown are discussed in the text.}
  \label{fig:thin-wire_schematic}
\end{figure}

% [JMM: I am not sure if this is actually worked out in the Kraus book]
For two closely spaced nanostructures, the $|\textbf{E}|^2$ enhancements in the resulting gap can in principle be explained using antenna theory \cite{Kraus_antenna-bible, Pohl_antenna-theory_for-NP_2000}, where the open-circuit voltage across the gap is responsible, and thus the systems are often classified as such \cite{bowtie_antennas_Moerner-PRL2005, opt-antennas_OJFMartin_Science-2005}. We therefore begin by considering a two-dimensional (2D) antenna as shown schematically in Fig.\ \ref{fig:thin-wire_schematic}.
[The extension to three dimensions (3D) will be discussed below.] Two metal wires M with widths $w$ are separated by a distance $a$ and the entire structure spans a length of $h$. The structure is illuminated from below at normal incidence by a plane wave with wavelength $\lambda$, and we wish to determine how the $|\textbf{E}|^2$ enhancement at the center of the gap depends on $a$. It is important to realize that for a real metal and distances less than approximately $1$ nm, nonlocal dielectric effects will become important \cite{McMahon_NLDiel}. Our quantitative analysis herein will therefore be for $a \geq 1$ nm, but in most cases we will include smaller distances to highlight qualitative features. Antenna theory for a perfect electrically conducting (PEC) thin-wire antenna ($w \ll h$) assumes that the incident electric field $\textbf{E}_0$ generates an alternating current along $x$, which results in an induced voltage $V$ across $h$. If $h \approx n \lambda / 2$, where $n$ is an integer, the antenna resonates and $V \approx -E_0 h$, where $E_0$ is the amplitude of the incident field \cite{Kraus_antenna-bible}. As a result, the open-circuit voltage in the gap should produce a uniform $|\textbf{E}|^2$ enhancement of $|\textbf{E}|^2 / |\textbf{E}_0|^2 = | V / a |^2 / | E_0 |^2 \approx h^2/a^2$.

\begin{figure}
  \includegraphics[scale=0.35, bb=0 0 685 537]{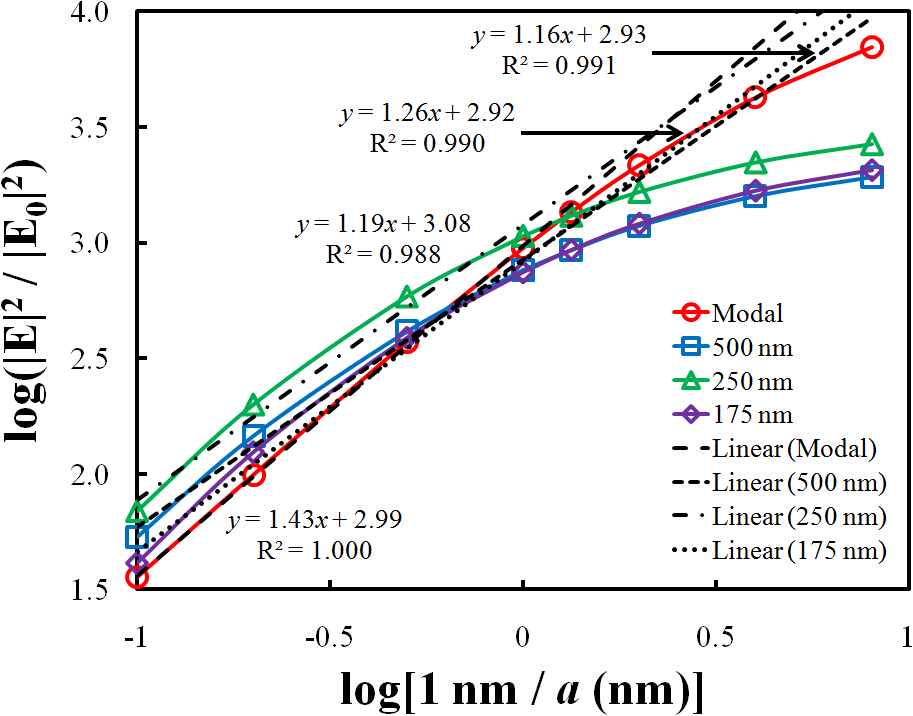}
  % 2D_structs_USE.png: 913x716 pixel, 96dpi, 24.16x18.95 cm, bb=0 0 685 537
  \caption{$|\textbf{E}|^2$ enhancements as a function of gap size for PEC thin-wire antennas with parameters given in the text. Solid lines are used to connect the actual data points (symbols).}
  \label{fig:2D_structs}
\end{figure}

In order to test the above analysis, we rigorously determined the $|\textbf{E}|^2$ enhancements via FEM calculations \cite{Jin_FEM} for $h = 250$ and $500$ nm thin-wire antennas ($n = 1$ and $2$, respectively) with $w = 5$ nm at $\lambda = 500$ nm for gap sizes of $a = 0.125$ to $10$ nm. 
%(Note that the off-resonance condition of $h = 175$ nm as well as a modal analysis calculation are also included, which we will discuss below.) 
To characterize the $a$ dependence, we can assume that $|\textbf{E}|^2 / |\textbf{E}_0|^2$ is proportional $1 / a^p$ and plot the results on a log-log scale to determine $p$; Fig.\ \ref{fig:2D_structs}.
For both antennas it is found that $p \approx 1.2$ for $a \geq 1$ nm, and it is even less for smaller $a$, which is much lower than the above antenna theory prediction of $p = 2$. 
% [JMM !] These results are quite different from the above antenna theory prediction of $p = 2$, which is surprising considering that these structures perfectly satisfies the definition of a thin-wire antenna. 
% [JMM] In addition to the resonance condition
% [JMM: antenna theory spot:] We will briefly discuss this below.
An alternative way to describe these systems and the $|\textbf{E}|^2$ enhancements that they exhibit is therefore needed.

The system in Fig.\ \ref{fig:thin-wire_schematic} can be greatly simplified by taking $h \rightarrow \infty$, which below we will show does not significantly affect the behavior of the electric field $\textbf{E}$ in the gap. Maxwell's equations can be solved analytically for such a system (if the metal is a PEC), which is an isolated slit in a metal film, by appropriately expanding the transverse component of the field (the $y$ component of the magnetic field, in this case) above and below the film and inside of the gap in terms of known functions and applying boundary conditions at the interfaces. While the full solution for this problem has been implicitly worked out in terms of a system of linear equations \cite{modal-analysis_FJGV_2004}, we demonstrate that under a few reasonable approximations it is possible to obtain a tractable semianalytical form for $|\textbf{E}|^2 / |\textbf{E}_0|^2$.

Inside the gap $\textbf{E}$ can be defined entirely in terms of its $x$ component $E_x$ (below we will show that the $z$ component $E_z$ is zero), which can be expanded as a superposition of forward and backwards propagating (and evanescent) waveguide modes $m$,
\begin{equation}
  \label{eq:Ex_all-m}
  E_x (x, z) = \sum_{m = 0}^\infty \frac{\beta_m}{k_0} \left( A_m e^{i \beta_m z} - B_m e^{-i \beta_m z} \right) \phi_m (x) ~~ ,
\end{equation}
%
% %
% \begin{equation}
%   \label{eq:Ez_all-m}
%   E_z (\xvec) = \frac{i}{k_0 \varepsilon} \sum_{m = 0}^\infty \left[A_m e^{i \beta_m z} + B_m e^{-i \beta_m z} \right] \varphi_m (x)
% \end{equation}
% %
%$\varphi_m (x) = - \left( 2 / a^{1/2} \right) (m \pi / a) \sin \left[ m \pi / a \left( x + a/2  \right) \right]$
where $A_m$ and $B_m$ are the respective modal amplitudes, $k_0 = 2 \pi / \lambda$, $\beta_m = \left[ k_0^2 - \left( m \pi / a \right)^2 \right]^{1/2}$, and $\phi_m (x) = \left( 2 / a^{1/2} \right) \cos \left[ m \pi \left( x + a/2  \right) / a \right]$ is the solution to the Helmholtz equation subject to PEC boundary conditions on the gap sides at $x = \pm a / 2$. $A_m$ and $B_m$ can be found by ensuring the continuity of Eq.\ (\ref{eq:Ex_all-m}) at the input ($I$) and output ($O$) surfaces of the gap,
\begin{equation}
  \label{eq:A_m}
  A_m = \frac{1}{2} \frac{k_0}{\beta_m} E_m^I \bigg \lbrace 1 + i \left[ \frac{1}{\tan (\beta_m w)} - \frac{E_m^O / E_m^I}{\sin (\beta_m w)} \right] \bigg \rbrace
\end{equation}
\begin{equation}
  \label{eq:B_m}
  B_m = \frac{1}{2} \frac{k_0}{\beta_m} E_m^I \bigg \lbrace -1 + i \left[ \frac{1}{\tan (\beta_m w)} - \frac{E_m^O / E_m^I}{\sin (\beta_m w)} \right] \bigg \rbrace
\end{equation}
where $E_m^I$ and $E_m^O$ are intensities of scattering events that take place at each of the surfaces, which can be determined by solving a set of linear equations (see below) \cite{modal-analysis_FJGV_2004}.

% [JMM: not sure if Jackson is best to ref here]
Inside a gap that is small relative to the incident wavelength ($k_0 a \ll 1$), the only waveguide mode that can exist is the $m = 0$ transverse electromagnetic (TEM) one \cite{Jackson_electrodynamics}. In this case, there is no $x$ dependence in $E_x$, $E_z$ is zero, $\beta_0 = k_0$, and the linear equations defining $E_0^I$ and $E_0^O$ simplify considerably,
\begin{equation}
  \label{eq:EI_0}
  E_0^I = I_0 \frac{ f_0 + g_{00} }{ \left( f_0 + g_{00} \right)^2 - \left( g_0^v \right)^2 }
\end{equation}
\begin{equation}
  \label{eq:EO_0}
  E_0^O = - \frac{g_0^v}{f_0 + g_{00}} E_0^I
\end{equation}
where $I_0 = 4 E_0 a^{1/2}$ (for normal incident light) is the overlap amplitude of the incident field with the TEM mode, $f_0 = i \cot \left( k_0 w \right)$ is the admittance amplitude, $g_0^v = i \csc \left( k_0 w \right)$ is the coupling amplitude between $I$ and $O$, and
\begin{equation}
  \label{eq:g_00_exact}
  g_{0 0} = \frac{4}{a} \int_{-a/2}^{a/2} \int_{-a/2}^{a/2} dx ~ dx' ~ G \left( x, x' \right)
\end{equation}
is the amplitude of the TEM mode's self-interaction, where $G \left( x, x' \right) = (k_0 / 2) H_0^{(1)} \left( k_0 | x - x'  | \right)$ is the 2D vacuum Green's function with $H_0^{(1)}$ being a Hankel function of the first kind. The maximum value that $k_0 | x - x' |$ can take is $k_0 a$. Since $k_0 a \ll 1$, we can make the small-argument approximation in $H_0^{(1)}$ and perform the integral in Eq.\ (\ref{eq:g_00_exact}) analytically,
\begin{equation}
  \label{eq:g_00_approx}
  g_{00} = 2 k_0 a \left( 1 + i l \right)
\end{equation}
where $l = \left( 2 / \pi \right) \left[ \ln \left( k_0 a / 2 \right) + \gamma - 3 / 2 \right]$ with $\gamma$ being Euler's constant.

For a gap with a small width relative to the incident wavelength ($k_0 w \ll 1$) we can use the small-angle approximation 
%[$\sin (x) \approx \tan (x) \approx x$ and $\cos (x) \approx 1$] 
in Eqs.\ (\ref{eq:A_m}) and (\ref{eq:B_m}) to greatly simplify Eq.\ (\ref{eq:Ex_all-m}),
%
% % [JMM: twocolumn]
% \begin{multline}
%    \label{eq:Ex_approx_full}
%    E_x (z) = E_0 8 \left[ \frac{ f_0 + g_{00} }{ \left( f_0 + g_{00} \right)^2 - \left( g_0^v \right)^2 } \right] \times \\
%    \left[ 1 - \frac{z}{w} \left( 1 - \frac{f_0}{f_0 + g_{00}} \right) \right] ~~ .
% \end{multline}
% %
%
\begin{equation}
  \label{eq:Ex_approx_full}
  E_x (z) = 8 E_0 \frac{ f_0 + g_{00} }{ \left( f_0 + g_{00} \right)^2 - \left( g_0^v \right)^2 } \left[ 1 - \frac{z}{w} \left( 1 - \frac{f_0}{f_0 + g_{00}} \right) \right] ~~ .
\end{equation}
Note that the approximation $g_{00} \ll 2 f_0$ was also used to get Eq.\ (\ref{eq:Ex_approx_full}), which is valid considering that the leading terms are $k_0 a$ and $1 / k_0 w$, respectively. 

At the center of the gap ($z = w / 2$) Eq.\ (\ref{eq:Ex_approx_full}) simplifies even further,
\begin{equation}
  \label{eq:Ex-approx_center-of-slit}
  E_x \left( w / 2 \right) = 8 E_0 \frac{ 1 }{ 2 g_{00} + u }
\end{equation}
where $u = f_0 - \left( g_0^v \right)^2 / f_0 = - i \tan \left( k_0 w \right) \approx - i k_0 w$. Equation (\ref{eq:Ex-approx_center-of-slit}) shows that at the center of the gap $\textbf{E}$ is inversely proportional to the interference between two terms, the self-interaction term $g_{00}$, which depends only on $k_0 a$, and a term $u$ representing the interference between surfaces $I$ and $O$, which depends only on $k_0 w$. 

Using the explicit expressions for $g_{00}$ and $u$ in Eq.\ (\ref{eq:Ex-approx_center-of-slit}) and calculating $|\textbf{E}|^2 / |\textbf{E}_0|^2 $ gives
%
% \begin{multline}
%   \label{eq:Ex-approx_terms-of-a}
%   | \textbf{E} |^2 / |\textbf{E}_0|^2 = | E_x \left( W / 2 \right) |^2 /|E_0|^2 = \\
%   \frac{64}{k_0^2} \left[ \frac{ 1 }{ 16 \left( 1 + l^2 \right) a^2 + 8 l a W + W^2 } \right] ~~ ,
% \end{multline}
%
% % [JMM: twocolumn JMM]
%  \begin{multline}
%    \label{eq:Ex-approx_terms-of-a}
%    | \textbf{E} |^2 / |\textbf{E}_0|^2 = | E_x \left( W / 2 \right) |^2 /|E_0|^2 = \\
%    \frac{ 64 }{ 16 \left( 1 + l^2 \right) \left( k_0 a \right)^2 + 8 l \left( k_0 a \right) \left( k_0 W \right) + \left( k_0 W \right)^2 } ~~ .
%  \end{multline}
% %
%
\begin{equation}
  \label{eq:Ex-approx_terms-of-a}
  | \textbf{E} |^2 / |\textbf{E}_0|^2 = | E_x \left( w / 2 \right) |^2 /|E_0|^2 = \frac{ 64 }{ 16 \left( 1 + l^2 \right) \left( k_0 a \right)^2 + 8 l \left( k_0 a \right) \left( k_0 w \right) + \left( k_0 w \right)^2 } ~~ .
\end{equation}
For gaps with $a$ not very small relative to $w$ (greater than approximately $1$ nm for the thin-wire antennas discussed above), $\textbf{E}$ depends primarily on $g_{00}$ and $| \textbf{E} |^2 / | \textbf{E}_0 |^2  \approx 64 / \left[ 16 \left( 1 + l^2 \right) \left( k_0 a \right)^2 \right]$. Because the enhancement is of the form $1 / l^2 a^2$, we would expect a $1 / a^p$ fit to not work at all. However, over a couple order of magnitude range of $a$ values $l \sim \ln \left( k_0 a / 2 \right)$ is well-approximated by $A \left( k_0 a / 2 \right)^b$, where $A$ and $b$ are constants that can be determined by demanding that this equality and a corresponding one involving its derivative be satisfied at some value of $k_0 a / 2$. With these constraints one finds that $b = 1 / \ln \left( k_0 a / 2 \right)$, which varies slowly with $k_0 a / 2$. For $\lambda = 500$ nm and $1 \leq a \leq 10$ nm, $b \approx -0.25$ and therefore $1 / l^2 a^2 \sim 1 / a^{1.5}$, which is consistent with $p \approx 1.2$ found for the thin-wire antennas in Fig. \ref{fig:2D_structs}. For small gaps both $a^2$ and $l^2 a^2$ go to zero, which means that $\textbf{E}$ depends primarily on $u$ and $| \textbf{E} |^2 / | \textbf{E}_0 |^2 \approx 64 / \left(  k_0 w \right)^2$. Therefore, for $a \ll w$ we expect a turnover to a weaker $1 / a^p$ dependence, which is also consistent with the results in Fig.\ \ref{fig:2D_structs}. Actual $|\textbf{E}|^2$ enhancements calculated using Eq.\ (\ref{eq:Ex-approx_terms-of-a}) for $a = 0.125$ to $10$ nm, $w = 5$ nm, and $\lambda = 500$ nm are shown in Fig.\ \ref{fig:2D_structs} as well, and are in agreement with these remarks. 

The strong agreement between the modal and thin-wire antenna results in Fig.\ \ref{fig:2D_structs} suggests that the $|\textbf{E}|^2$ enhancements in both cases arise from the same effect, a TEM waveguide mode. The former does show a slightly stronger $1 / a^p$ dependence, but this can be understood as follows. Recall that the amplitude for coupling incident light into this mode is proportional to their overlap; see Eq.\ (\ref{eq:EI_0}). In a finite structure, such as a thin-wire antenna, some of the impinging incident light can be effectively lost via scattering, leading to a less efficient coupling into the TEM mode and a weaker $1 / a^p$ dependence. Quantitatively, the amount of scattering is given by the scattering efficiency $Q_\text{sc}$ (the ratio of the scattering cross section to the geometric one). In the modal results $Q_\text{sc}$ is naturally $0$, since the geometric cross section is infinite. For a finite structure, however, $Q_\text{sc} \geq 0$. For example, the resonant antennas $h = 500$ and $250$ nm have similar $Q_\text{sc}$ values of $1.953$ and $1.915$, respectively, at $a = 2$ nm. For an off-resonance condition we expect less scattering, and in fact this is what is numerically found for $h = 175$ nm. In this case, $Q_\text{sc} = 0.931$ nm (at $a = 2$ nm) and the $1 / a^p$ dependence is indeed stronger than for the two resonant antennas.
% (Note that $Q_\text{sc}$ is much higher for the resonance conditions.)

%[JMM: ?? antenna theory spot: perhaps the reconciliation with antenna theory should be made here -- the text doc on the laptop kind of explains this]

Thusfar we have considered 2D systems. While such structures are experimentally realizable, in most cases (e.g., typical SERS substrates) structures are 3D in character. Nonetheless, our analysis remains valid and Eq.\ (\ref{eq:Ex-approx_terms-of-a}) should still apply (qualitatively, at least). This is because the TEM waveguide mode suggested as responsible for the $|\textbf{E}|^2$ enhancements also exists in 3D. The only requirement to support a propagating electromagnetic wave is that an oscillating potential difference be established between the walls supporting the wave (e.g., the sides of a gap) \cite{SlaterFrank_electromagnetism}. In addition to 3D, actual structures are comprised of a real metal, typically Ag or Au at optical frequencies due to possible increases in $|\textbf{E}|^2$ enhancements via surface plasmon (SP) excitations \cite{EM_fields_NP-dimers_GCS, EM_fields_dimers_Martin-2001}. Based on the discussion above about $Q_\text{sc}$, it is reasonable to suspect that the dominant effect of this is that there may be wavelength dependent modulations to the $1 / a^p$ dependence, due to corresponding dependencies in the absorption efficiency ($Q_\text{abs}$) and $Q_\text{sc}$ \cite{Light-Scattering_BH}. However, the overall trends and underlying physical principles should remain the same.

\begin{figure}
  \includegraphics[scale=0.35, bb=0 0 693 537]{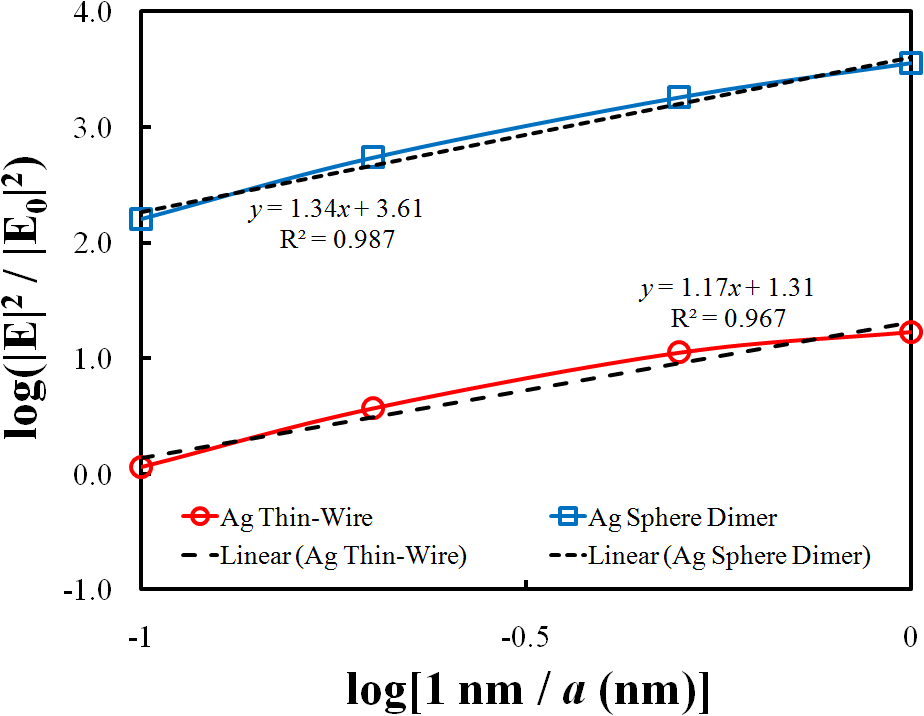}
  % 3D_structs_USE.png: 924x716 pixel, 96dpi, 24.45x18.95 cm, bb=0 0 693 537
  \caption{$|\textbf{E}|^2$ enhancements as a function of gap size for a 3D Ag thin-wire antenna and a dimer of Ag spheres with parameters given in the text. Solid lines are used to connect the actual data points (symbols).}
  \label{fig:3D_structs}
\end{figure}

As a first example of the applicability of our analysis to real 3D structures, FEM was used to calculate the $1 / a^p$ dependence at the center of a 3D Ag thin-wire antenna with $h = 250$ nm and $w = 5$ nm (both in and out of the plane of Fig.\ \ref{fig:thin-wire_schematic}) at $\lambda = 500$ nm; Fig.\ \ref{fig:3D_structs}.
%
%(Note that at $\lambda = 500$ nm $\varepsilon_\text{Ag} = -9.8 + i 0.32$ \cite{OptConst_NobleMetals_JC}, and in this case the $|\textbf{E}|^2$ enhancements are actually not higher than for the PEC). 
It is found that $p$ is again approximately $1.2$, which is nearly equal to the analogous 2D PEC thin-wire antenna results in Fig.\ \ref{fig:2D_structs}. Further similarity comes from the behavior as $a$ decreases, where the $1 / a^p$ dependence again becomes weaker, as can be inferred from the curvature of the actual data relative to the linear fit.
% [JMM] In passing, we should mention that while the results are in good agreement here, it is possible that local deviations from the trends can be observed, considering that the absorption efficiency of a plasmonic structure has both associated resonance peaks and interferences that produce valleys \cite{McMahon_2010_JACS_SERS-trimers, EM_fields_dimers_Martin-2001}. However, the overall trends should be consistent.
% [JMM: rev b comment 1]

\begin{figure}
  \includegraphics[scale=0.42, bb=0 0 573 494]{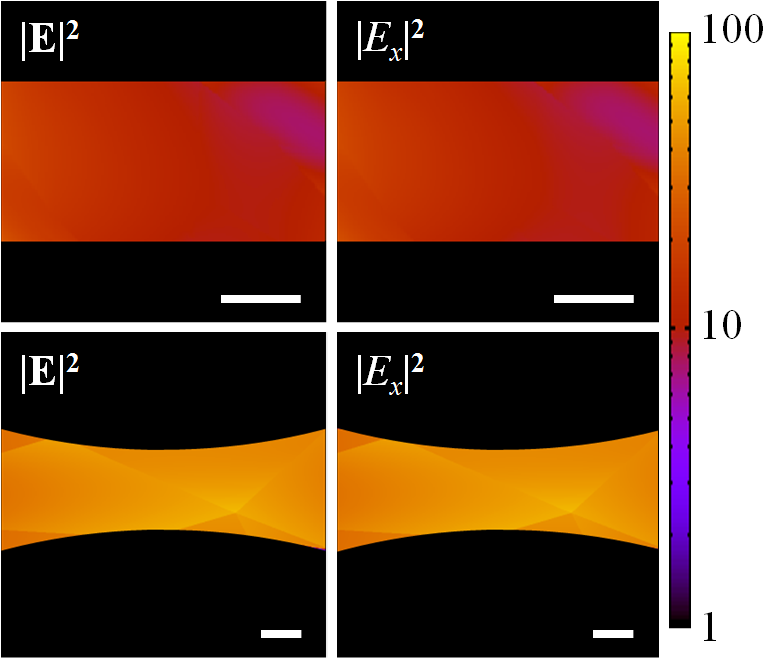}
  % Ag_field-prof_USE.png: 764x659 pixel, 96dpi, 20.22x17.44 cm, bb=0 0 573 494
  \caption{Intensities of $|\textbf{E}|^2$ and the incident component of $\textbf{E}$, $|E_x|^2$, in the gaps of a 3D Ag thin-wire antenna (top) and Ag sphere dimer (bottom) with parameters discussed in the text. The inset white scale bars correspond to $1$ nm. Note that the intensity values have been rescaled relative to Fig.\ \ref{fig:3D_structs} to fit clearly on the same scale, and the fields inside of the structures have been set to $0$.}
  \label{fig:field_prof}
\end{figure}

It is possible to verify the existence of a TEM waveguide mode by looking at profiles of $\textbf{E}$ inside the gap, as for normal incident linearly-polarized light this mode has the same polarization. Figure \ref{fig:field_prof} shows the fields inside the gap of the antenna discussed above for $a = 2$ nm, and it can be seen that this is indeed the case, as $|\textbf{E}|^2 \approx |E_x|^2$. 
(Note that $|E_y|^2$ and $|E_z|^2$ are both less than $1$ on the scale in Fig.\ \ref{fig:field_prof}.) In fact, it has recently been demonstrated experimentally that electromagnetic fields inside the gaps of nanostructures are linearly polarized, even in more complex ones than discussed here \cite{gap-fields_lin-pol_Aizpurua-NanoLett2010}.

As a further and final example of the applicability of our analysis to real 3D structures, $|\textbf{E}|^2$ enhancements as a function of gap size were calculated for a dimer of $250$ nm diameter Ag spheres at $\lambda = 633$ nm (a popular type of experimental system and common laser wavelength \cite{McMahon_2010_JACS_SERS-trimers}); Fig.\ \ref{fig:3D_structs}. The $1 / a^p$ dependence in this case is found to be characterized by $p \approx 1.3$, which is nearly equal to the modal result of $p \approx 1.4$, and slightly greater than the somewhat analogous thin-wire antenna result in Fig.\ \ref{fig:3D_structs}. The stronger $1 / a^p$ dependence is related to the fact that this structure is more efficient for capturing light \cite{Light-Scattering_BH}, as indicated by $Q_\text{abs} = 0.200$ as opposed to $0.087$ for the thin-wire antenna, for example.
% [Q_sc = 5.942 and 0.004 respectively]
% [JMM: the comment below about cross sections may not be correct. You could say absorption cross sections,
% but for the PEC this has to be 0, so it would be related to scattering seems to be in some sort of conflict
% with other results.]
%(this can be seen by comparison of extinction cross sections, for example -- not shown). 
% [JMM: here are scattering & absorption numbers (5 nm separation)
% Fig. 2: PEC = s:975, Ag = s:111 a:25
% Fig. 2: PEC = s:676, Ag = s:1173 a:42
% This is also reflected in Fig.\ \ref{fig:FEM_cyl_calcs}, where $p \approx 2.2$ for $a \geq 1$ nm in the Ag dimer results.
% [JMM: !! it is interesting that ]
Field profiles inside the gap again indicate the presence of a TEM waveguide mode; Fig.\ \ref{fig:field_prof}. Looking closely at Fig.\ \ref{fig:3D_structs} reveals that there is a much less strong turnover to a weaker $1 / a^p$ dependence for smaller $a$ than was seen for any of the other structures. Such behavior is understandable considering that $w$ is effectively zero in this case (there is only a single point of minimum approach), which can lead to $| \textbf{E} |^2$ being unbounded as $a \rightarrow 0$ [$\lim_{a \to 0} \left( 1 + l^2 \right) \left( k_0 a \right)^2 = 0$; see Eq.\ (\ref{eq:Ex-approx_terms-of-a})]. %These results indicate that the highest $| \textbf{E} |^2$ enhancements for small $a$ will occur for structures with $W$ effectively zero.
It is quite remarkable that the simple analysis derived for a 2D PEC film with an isolated slit is so accurate when applied to full 3D structures of other geometries, and even in the presence of SPs.

% [JMM !!!] Through FEM calculations and semianalytic theory, we demonstrated that the $|\textbf{E}|^2$ enhancement in the gap between any two nanostructures, irrespective of the material or geometry, is the result of a TEM waveguide mode. The results presented showed that the enhancement depends on the gap size in a way that can be characterized by $1 / a^p$, where $p$ is related to the efficiency of the nanostructure to couple incident light into the TEM mode. In the case of an isolated slit in a PEC film, which we used as a reference system, $p \approx 1.4$. In a system which exhibits additional scattering, such as a PEC thin-wire antenna, $p < 1.4$ and in extreme cases can be lower than $1$. On the other hand, in plasmonic structures that can efficiently capture the incident light, $p > 1.4$ and in some cases can be greater than $2$. The results presented are important for a fundamental understanding of the $|\textbf{E}|^2$ enhancements that occur in the gaps between closely spaced nanostructures.

\begin{acknowledgments}
J.M.M. and G.C.S. were supported by a Grant from the U. S. Department of Energy, Office of Science, Office of Basic Energy Sciences, under Award No.\ DE-SC0004752. Use of the Center for Nanoscale Materials was supported by the \text{U. S.} Department of Energy, Office of Science, Office of Basic Energy Sciences, under Contract No.\ DE-AC02-06CH11357.
\end{acknowledgments}

%\bibliography{../Bibtex_Refs/journal_names_s,../Bibtex_Refs/SERS_related,../Bibtex_Refs/EM_enhancements,../Bibtex_Refs/EM-methods,../Bibtex_Refs/McMahon_pubs,../Bibtex_Refs/diel-modeling,../Bibtex_Refs/nanoparticles}

\end{document}